\documentclass[a4paper,12pt,abstract=true]{scrartcl}
\usepackage{authblk}
\usepackage{amsmath,amssymb}
\usepackage{bm}
\usepackage{graphicx,color}
\usepackage[hidelinks]{hyperref}

\usepackage[all,warning]{onlyamsmath}
\RequirePackage[l2tabu, orthodox]{nag}

\newbox{\ORCIDicon}
\sbox{\ORCIDicon}{\large
                  \includegraphics[width=0.8em]{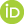}}

\begin{document}

\titlehead{\hfill OU-HET 1029}

\title{Relativistic effects in search for new intra-atomic force
       with isotope shifts}

\author[1]{Minoru~Tanaka\,\href{https://orcid.org/0000-0001-8190-2863}
                               {\usebox{\ORCIDicon}}
           \thanks{Email: \texttt{tanaka@phys.sci.osaka-u.ac.jp}}}

\affil[1]{Department of Physics, Graduate School of Science, 
          Osaka University, Toyonaka, Osaka 560-0043, Japan} 

\author[2]{Yasuhiro~Yamamoto
           \thanks{Email: \texttt{ph.yamayasu@gmail.com}}}
\affil[2]{National Centre for Nuclear Research, 
          Pasteura 7, 02-093 Warsaw, Poland}

\date{\normalsize\today}

\maketitle

\begin{abstract}
Isotope shift of atomic spectra is considered as a probe of 
new interaction between electrons and neutrons in atoms.
We employ the method of seeking a breakdown of King's linearity
in the isotope shifts of two atomic transitions.
In the present work, we evaluate the magnitudes of the nonlinearity
using relativistic wave functions and the result is compared with 
that of nonrelativistic wave functions in our previous work.
It turns out that the nonrelativistic calculation underestimates 
the nonlinearity owing to the new interaction in the mass range
of the mediator greater than 1 MeV.
Further, we find that the nonlinearity within the standard model
of particle physics is significantly magnified by 
the relativistic effect in the $\text{p}_{1/2}$ state.
To get rid of this obstacle in the new physics search,
we suggest to avoid $\text{p}_{1/2}$, and use $\text{p}_{3/2}$ 
instead for example.
\end{abstract}

\newpage
\section{\label{Sec:Intro}Introduction}
In developing the next-generation frequency standard, the precision 
of atomic spectroscopy has been considerably improved. For instance,
the relative uncertainty of $O(10^{-18})$ is realized in a clock 
transition of $\text{Yb}^+$~\cite{Huntemann2016a}.
The corresponding accuracy of frequency (energy) is of the order of mHz
($10^{-18}$ eV). 

We consider a possibility to constrain or probe a new physics beyond 
the standard model of particle physics with such high precision 
spectroscopy. It is practically impossible to calculate transition 
frequencies of many-electron atomic systems at the level of the above
accuracy. To overcome this difficulty, a method to use isotope
shifts (IS) was proposed~\cite{DelaunayOzeriPerezSoreq2016a}
and further studied in a quantitative manner~\cite{Berengut2017a}.%
\footnote{For one- and two-electron atoms, constraints 
on new physics using their IS are studied in 
Ref.~\cite{DelaunayFrugiueleFuchsSoreq2017a}.}

The typical magnitude of observed IS is the order of GHz 
($10^{-6}$ eV), much smaller than optical transition energies,
and such quantities are evaluated by perturbation.
If nuclei were infinitely heavy and point-like, the electrons in an atom
could not sense the difference of the nuclei among isotopes and
no IS would exist.
Thus, it is customary to decompose observed isotope 
shifts into the mass shift (MS), which is due to the finite nuclear mass,
and the field shift (FS), due to the finite nuclear size~\cite{King1984a}.
In other words, the nuclear recoil causes the MS and
the change of the nuclear Coulomb field does the FS.

The isotope shift of a transition (labeled by $t$) between an isotope
pair $A$ and $A'$, $\nu_{t,A'A}:=\nu_{t,A'}-\nu_{t,A}$, in the leading 
order perturbation is expressed by
\begin{equation}\label{Eq:IS}
\nu_{t,A'A}=K_t\mu_{A'A}+F_t\langle r^2\rangle_{A'A}\,,
\end{equation}
where $\mu_{A'A}:=\mu_{A'}-\mu_A$ represents the difference of 
the reduced masses, 
$\langle r^2\rangle_{A'A}:=\langle r^2\rangle_{A'}-\langle r^2\rangle_{A}$ 
denotes that of the mean-squared radii of nuclei, 
and $K_t$ and $F_t$ are electronic factors. 
The electronic factors solely depend on the transition 
and are common for isotopes.

The key idea to explore new physics using IS is King's linearity%
~\cite{King1963a} and its breakdown by a new interaction%
~\cite{DelaunayOzeriPerezSoreq2016a}.
Suppose isotope shifts of several isotope pairs in two distinct 
transitions ($t=1,2$) of an atom (or an ion).
It is convenient to introduce the modified isotope shift (mIS)
$\tilde\nu_{t,A'A}:=\nu_{t,A'A}/\mu_{A'A}$ so that Eq.~\eqref{Eq:IS}
is written as
\begin{equation}\label{Eq:mIS}
\tilde\nu_{t,A'A}=K_t+F_t\frac{\langle r^2\rangle_{A'A}}{\mu_{A'A}}\,.
\end{equation}
We note that atomic masses are measured 
with typical precision of $10^{-8}$ or better~\cite{Wang2012a}.%
\footnote{In our numerical calculation, we use 
the relevant atomic masses given in Ref.~\cite{Wang2012a} to compute 
the reduced masses. 
The difference between atomic and nuclear masses cancels in $\mu_{A'A}$
at the leading order. The remaining correction is higher order 
in the MS. } 
Eliminating the nuclear factor $\langle r^2\rangle_{A'A}/\mu_{A'A}$, 
we obtain a linear relation between the mIS's of two transitions,
\begin{equation}\label{Eq:Linearity}
\tilde\nu_{2,A'A}=K_{21}+F_{21}\tilde\nu_{1,A'A}\,,
\end{equation}
where $K_{21}:=K_2-F_{21}K_1$ and $F_{21}:=F_2/F_1$.

This linear relation is broken by a new force of particle exchange 
between an electron and a neutron. 
In the presence of such particle exchange, a particle shift (PS),
a new source of IS, is added to the IS expression in Eq.~\eqref{Eq:IS},
\begin{equation}\label{Eq:ISPS}
\nu_{t,A'A}=K_t\mu_{A'A}+F_t\langle r^2\rangle_{A'A}+X_t(A'-A)\,.
\end{equation}
The linear relation in Eq.~\eqref{Eq:Linearity} is also modified as
\begin{equation}\label{Eq:NL}
\tilde\nu_{2,A'A}=K_{21}+F_{21}\tilde\nu_{1,A'A}
                  +\varepsilon_\text{PS}A'A\,,
\end{equation}
where
\begin{equation}\label{Eq:PSNL}
\varepsilon_\text{PS}:=\frac{M}{m_e^2}(X_2-F_{21}X_1)\,,
\end{equation}
and $M$ denotes the atomic mass unit. 
We use $\varepsilon_\text{PS}$ to quantify the nonlinearity due to PS
for a pair of transitions.
We note that values of $\varepsilon_\text{PS}$ for different pairs of
transitions cannot be directly compared in general.

In the standard model, as pointed out in 
Ref.~\cite{DelaunayOzeriPerezSoreq2016a}, Z or Higgs boson leads to such
a nonlinearity. 
However, the nonlinearities by such heavy particles are strongly 
suppressed and its measurement is beyond the reach of the present technology%
~\cite{DelaunayOzeriPerezSoreq2016a,MikamiTanakaYamamoto2017}.
If there exists a light boson that couples to the electron and the neutron,
the nonlinearity may be detectable depending on its mass and couplings. 

Other conceivable sources of nonlinearity within the standard model
are discussed in the literature.
It is pointed out in Ref.~\cite{Blundell1987a} that
the subleading FS's in the $\text{s}_{1/2}$ and $\text{p}_{1/2}$ states
are likely to cause a nonlinearity as well as the FS in the second 
order perturbation. 
A rough estimate of nonlinearities is given in 
Ref.~\cite{FlambaumGeddesViatkina2017a}, in which approximate
analytic wave functions are used. 

The effect of the subleading FS other than the leading contribution
shown in Eq.~\eqref{Eq:IS} (e.g.~a term proportional to 
$\langle r^4\rangle_{A'A}$ as described in Sec.~\ref{Sec:NL}), is
effectively included in Eq.~\eqref{Eq:NL}
by replacing $\varepsilon_\text{PS}$ with 
$\varepsilon_\text{PS}+\varepsilon_\text{FS}$,
where $\varepsilon_\text{FS}$ represents the FS nonlinearity. 
The FS nonlinearity is a potential background in the search of new physics 
with the IS nonlinearity, though it can be removed by using three or
more transitions in principle~\cite{MikamiTanakaYamamoto2017}.

In this work, we evaluate the PS and FS nonlinearities employing 
numerical relativistic wave functions 
and compare the result with
that of our previous work with nonrelativistic (NR) wave functions%
~\cite{MikamiTanakaYamamoto2017}. We find that the sensitivity to 
a new electron-neutron interaction is enhanced in the mediator mass 
range of about 1 to 10 MeV in our relativistic evaluation but the FS 
nonlinearity significantly increases as far as 
a $\text{p}_{1/2}$ state is involved in the relevant transitions.

The rest of the paper is organized as follows.
We present the formulation to evaluate PS and FS in
Sec.~\ref{Sec:Formulation}.
In Sec. \ref{Sec:RWF}, our method to obtain relativistic wave functions
is described. 
We give formulas to estimate PS and FS nonlinearities with
the obtained wave functions in Sec.~\ref{Sec:NL}.
The current experimental status and future prospects are given in
Sec.~\ref{Sec:StatusProspect}.
Section~\ref{Sec:Conclusion} is devoted to our conclusion.
We use the natural units, $\hbar=c=1$, unless otherwise stated.

\section{\label{Sec:Formulation}Formulation of isotope shift}
As in the previous work, we employ the single-electron approximation to
evaluate PS and FS. One valence electron is supposed to change
its state (orbital) in a transition while the state of the other 
electrons is kept intact. 
We denote the radial densities of the initial and final states of 
this electron by $\sigma_i(r)$ and $\sigma_f(r)$ respectively.
We account for them in Sec.~\ref{Sec:RWF}.

A new spin-independent force between an electron and a neutron 
by exchanging a scalar or vector boson of mass $m$ is described
by the potential,
\begin{equation}\label{Eq:NewPot}
V(r)=(-1)^{s+1}\frac{g_n g_e}{4\pi}\frac{e^{-mr}}{r}\,,
\end{equation}
where $s$ is the mediator spin, and $g_n$ and $g_e$ denote
the neutron and electron coupling constants respectively.%
\footnote{In the present work, we consider isotopes of spin-0 nuclei,
so that the possible spin-dependent part of the new force is irrelevant.}
The PS electronic factor in Eq.~\eqref{Eq:ISPS} is given by
\begin{equation}\label{Eq:PS}
X_t=\int dr\,r^2 V(r)\sigma_t(r)\,,
\end{equation}
where $\sigma_t(r):=\sigma_i(r)-\sigma_f(r)$ and the radial densities are
normalized as $\int dr\,r^2\sigma_{i,f}(r)=1$. 
We note that the point-like nucleus is assumed in Eq.~\eqref{Eq:NewPot} 
so that it is valid for the inverse of the mediator mass
larger than the nuclear radius $\sim (100\ \text{MeV})^{-1}$.

We also evaluate the FS with the same electron densities to 
estimate the magnitude of the FS nonlinearity. 
The FS is the variation of the transition energy owing to 
the difference of the Coulomb potential of two nuclei $A$ and $A'$. 
Assuming spherically symmetric nuclear charge distributions, 
the potential difference between isotopes is also spherically symmetric
and represented by
\begin{equation}
V_{A'A}(r)=-Z\alpha\int d^3r'\,\frac{\rho_{A'A}(r')}{|\bm{r}-\bm{r}'|}\,,
\end{equation}
where $Z$ is the nuclear charge, $\rho_{A'A}(r):=\rho_{A'}(r)-\rho_{A}(r)$,
and the spherical nuclear charge distribution of an isotope $A$ denoted by
$\rho_{A}(r)$ is normalized as $4\pi\int dr\,r^2\rho_A(r)=1$. 
Then, the FS of transition $t$ is given by
\begin{equation}\label{Eq:FS}
\nu_{t,A'A}\bigl|_\text{FS}=\int dr\, r^2 V_{A'A}(r)\sigma_t(r)\,.
\end{equation}

\section{\label{Sec:RWF}Relativistic wave functions}
In order to evaluate the radial electron density $\sigma_{i,f}(r)$
in a relativistic manner, we solve the Dirac equation
with an effective potential that represents the potential by
the nucleus and the electrons other than the one involved
in a transition.

\subsection{Nuclear charge distribution}
Since the nuclear potential difference $V_{A'A}(r)$ vanishes outside
the nuclei, the FS in Eq.~\eqref{Eq:FS} is governed by the wave 
functions inside the nuclei. 
Hence it is important to evaluate the wave functions near the origin 
taking account of the finite nuclear size~\cite{Blundell1987a}.

We employ the Helm distribution~\cite{Helm1956a}, 
which is the gaussian-smeared uniform sphere,
\begin{equation}
\rho_A(r):=\int d^3r'\,
            \frac{3}{4\pi r_A^3}\theta(r_A-r')
            \frac{1}{(2\pi s^2)^{3/2}}e^{-|\bm{r}-\bm{r}'|^2/2s^2}.
\end{equation}
The nuclear Coulomb potential of the above Helm charge distribution
is given by~\footnote{There are typographical errors in Eq.~(B.12)
in Ref.~\cite{MikamiTanakaYamamoto2017}.} 
\begin{align}
V_A(r)=&-Z\alpha\int d^3r'\,\frac{\rho_{A}(r')}{|\bm{r}-\bm{r}'|}\\
      =&-\frac{Z\alpha}{4\pi r_A^3 r}
        \biggl[\sqrt{2\pi}s
               \biggl\{(r^2+r_A r-2r_A^2+2s^2)e^{-(r-r_A)^2/2s^2}\nonumber\\
       &              -(r^2-r_A r-2r_A^2+2s^2)e^{-(r+r_A)^2/2s^2}\biggr\}
        \nonumber\\
       &+\pi\biggl\{\left((r-r_A)^2(r+2r_A)+3s^2 r\right)
                   \text{Erf}\left(\frac{r-r_A}{\sqrt{2}s}\right)\nonumber\\
       &           -\left((r+r_A)^2(r-2r_A)+3s^2 r\right)
                    \text{Erf}\left(\frac{r+r_A}{\sqrt{2}s}\right)\biggr\}
        \biggr]\,,
\end{align}
where the error function is defined by
\begin{equation}
\text{Erf}(x):=\frac{2}{\sqrt{\pi}}\int^x_0 e^{-t^2}dt\,,
\end{equation}
so that  $\text{Erf}(+\infty)=1$.
A more detailed description of the Helm distribution is found in
Appendix B in Ref.~\cite{MikamiTanakaYamamoto2017}.
Following Ref.~\cite{LewinSmith1995a}, we use
$r_A^2=c_A^2+(7/3)\pi^2 a^2-5 s^2$ with
$c_A\simeq 1.23 A^{1/3}-0.60\ \text{fm}$, $a\simeq 0.52\ \text{fm}$
and $s\simeq 0.9\ \text{fm}$ in our numerical calculation.

\subsection{Effective potential}
The electronic states in the single-electron approximation are determined
by an effective potential that describes the nuclear field and 
the field of the other electrons kept unchanged 
in a transition.
As in our previous work, we make use of the Thomas-Fermi (TF) model, 
in which the ensemble of atomic electrons contributing to 
the effective potential is treated as a free Fermi gas in 
a slowly varying external potential.
It gives the following effective potential for a point-like nucleus,
\begin{equation}\label{Eq:TFP}
V_\text{TF}(r)=-\frac{Z\alpha}{r}\chi(r/b)
               -n\alpha\,\text{min}\left(\frac{1}{r_0},\frac{1}{r}\right)\,,
\end{equation}
where $n$ is the total charge (in the unit of $|e|$) of the system
described by the effective potential, $r_0=b x_0$ with $x_0$ 
the zero of the TF function $\chi(x)$, $b=[(3\pi)^2/(2^7 Z)]^{1/3}a_B$
and $a_B:=1/\alpha m_e$ is the Bohr radius. 

The TF function satisfies
\begin{equation}
\frac{d^2\chi}{dx^2}=\left\{\begin{matrix}
                             x^{-1/2}\chi^{3/2}\,, & \chi>0 \\
                             0\,,                  & \chi<0
                            \end{matrix}\right.
\end{equation}
with $\chi(0)=1$. The second boundary condition that uniquely specifies
the solution is $x_0\chi'(x_0)=-n/Z$ with $x_0$ being the solution of
$\chi(x_0)=0$ for a positive ion ($n\geq 1$). 
In the following part of the paper, we examine IS's of singly charged
positive ions so that the single-electron orbitals are determined by
the effective potential of $n=2$. See e.g.~Ref.~\cite{March1975a}
for details of the TF model.

The potential in Eq.~\eqref{Eq:TFP} is obtained for the case of
the point-like nucleus. 
As stressed above, the finite nuclear size should not be overlooked 
in the FS calculation.
Accordingly we modify the TF potential by subtracting the Coulomb
potential of the point-like nucleus, $V_c(r)=-Z\alpha/r$, 
and adding the potential of the Helm distribution $V_{A}(r)$,
\begin{equation}\label{Eq:mTFP}
V_\text{mTF}(r)=V_\text{TF}(r)-V_c(r)+V_A(r)\,.
\end{equation}
This modification significantly alters the behaviors of some wave
functions inside the nucleus though it affects the spectra little.

\subsection{Dirac equation}
The relativistic wave function of an electron bounded in the modified
TF potential in Eq.~\eqref{Eq:mTFP} is obtained by solving
the eigenvalue problem of the Dirac equation,
\begin{equation}
\left[-i\bm{\alpha}\cdot\bm{\nabla}+\beta m_e+V_\text{mTF}(r)
\right]\psi(\bm{r})=E\psi(\bm{r})\,,
\end{equation}
where $m_e$ is the electron mass, and $\bm{\alpha}$ and $\beta$ denote
the Dirac matrices. 
It is well known that the eigenfunction of this central-force problem
is expressed in the form of separation of variables in the spherical
coordinates as
\begin{equation}
\psi(\bm{r})=\begin{pmatrix}
              \dfrac{G(r)}{r}
               \mathcal{Y}^{j_3}_{j\ell_A}(\theta,\varphi)\\[2ex]
              i\dfrac{F(r)}{r}
               \mathcal{Y}^{j_3}_{j\ell_B}(\theta,\varphi)
             \end{pmatrix}\,,
\end{equation}
where $\mathcal{Y}^{j_3}_{j\ell}(\theta,\varphi)$ represents 
the spinor spherical harmonics, and $G(r)$ and $F(r)$ are radial
wave functions. See e.g.~Ref.~\cite{Sakurai1967a}.

The radial wave functions satisfy
\begin{align}
&\frac{dF}{dr}-\frac{\kappa}{r}F=-[E-V_\text{mTF}(r)-m_e]G\,,\label{Eq:RE1}\\
&\frac{dG}{dr}+\frac{\kappa}{r}G= [E-V_\text{mTF}(r)+m_e]F\,,\label{Eq:RE2}
\end{align}
where $\kappa=\pm(j+1/2)$. 
The orbital angular momenta of the spinor spherical harmonics are 
given as $(\ell_A,\ell_B)=(j+1/2,j-1/2)$ for $\kappa=j+1/2$ and 
$(\ell_A,\ell_B)=(j-1/2,j+1/2)$ for $\kappa=-(j+1/2)$.
The normalization of the wave function is 
$\int d^3r\,\psi^\dagger \psi=\int dr\,(G^2+F^2)=1$.
The radial electron density is defined in terms of the radial wave functions 
by $\sigma(r):=[G^2(r)+F^2(r)]/r^2$, so that $\int dr\,r^2\sigma(r)=1$
as mentioned below Eq.~\eqref{Eq:PS}.

\begin{figure}
 \centering
 \includegraphics[width=0.7\textwidth]{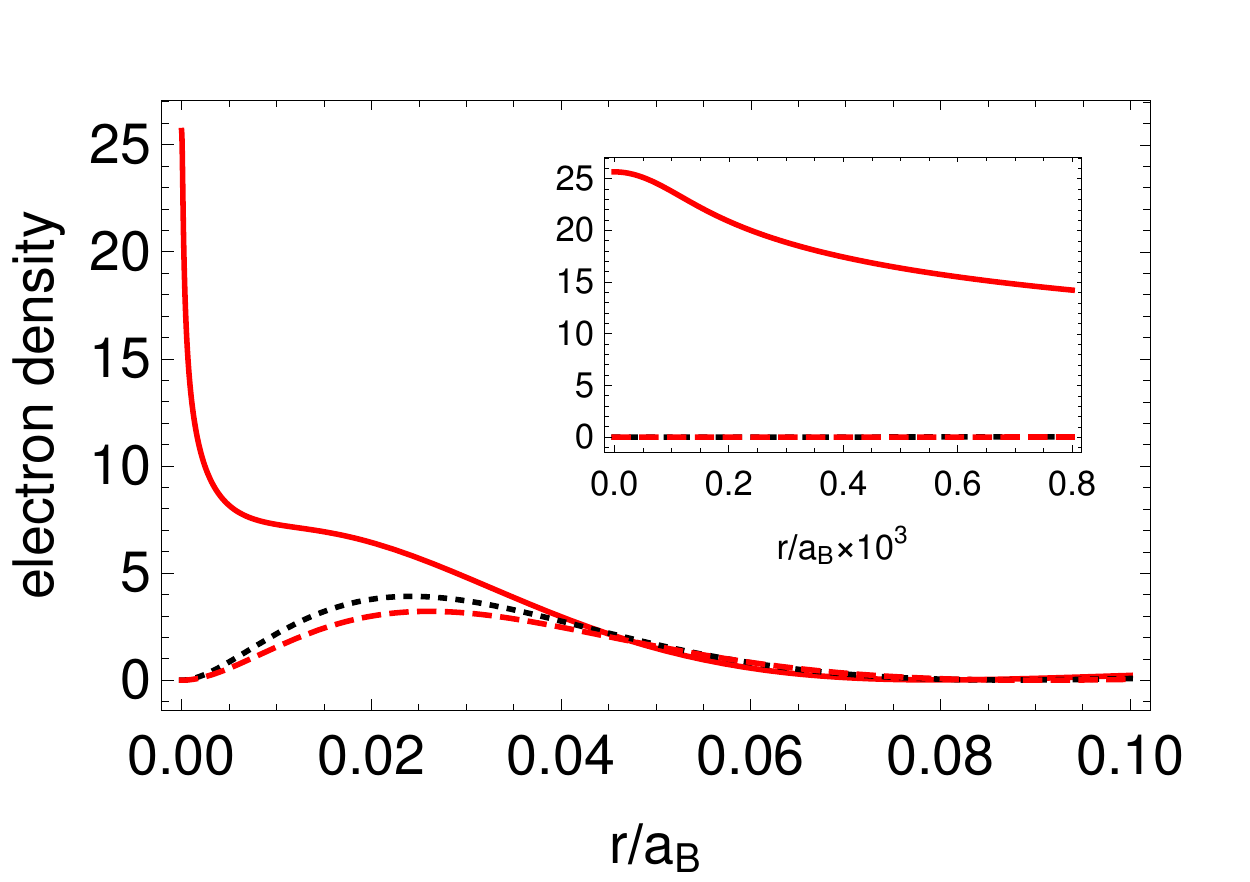}
 \caption{Illustration of wave functions. The electron density
          $\sigma(r)$ of the $\text{Yb}^+$ $6\text{p}_{1/2}$ state
          is plotted in the atomic units (solid red). 
          The NR $6\text{p}$ density (black dotted) is also
          presented for comparison. 
          We show $6\text{p}_{3/2}$ (red dashed) as well for
          a later discussion.
          The inset is a close look near the nucleus, whose size is 
          $1.2\times 10^{-4}a_B$.}
 \label{Fig:WF}
\end{figure}

We solve the radial equations in Eqs.~\eqref{Eq:RE1} and \eqref{Eq:RE2}
with the numerical method of shooting described in 
Ref.~\cite{SilbarGoldman2010a}.
Figure~\ref{Fig:WF} illustrates relativistic wave functions and
compares them to the NR one.
The behavior of the $\text{Yb}^+$ $6\text{p}_{1/2}$ state near
the origin is saliently different from the NR $6\text{p}$ state,
As discussed below, this difference results in an enhancement
of the FS nonlinearity in the relativistic calculation.
We note that the $6\text{p}_{3/2}$ behaves in a similar way as
the NR state and its possible benefit is discussed in
Sec.~\ref{Sec:StatusProspect}.

\section{\label{Sec:NL}Nonlinearity formulas}
The leading contribution in the FS, $F_t$ in 
Eq.~\eqref{Eq:IS}, 
is identified by expanding $\sigma_t(r)$ as
$\sigma_t(r)=\sigma_t(0)+\sigma''_t(0) r^2/2+\dotsb$.
We note that, as illustrated in the inset of Fig.~\ref{Fig:WF},
no linear term appears provided that the nuclear charge distribution
has no cusp at the origin~\cite{Blundell1987a}.

Rewriting Eq.~\eqref{Eq:FS} as
\begin{equation}\label{Eq:FS2}
\nu_{t,A'A}\bigl|_\text{FS}=
 -4\pi Z\alpha\int_0^\infty dr'\, r'^2\rho_{A'A}(r')\int_0^{r'}dr\, r^2
               \left(\frac{1}{r'}-\frac{1}{r}\right)\sigma_t(r)\,,
\end{equation}
it is straightforward to find that
\begin{equation}
\nu_{t,A'A}\bigl|_\text{FS}=
 Z\alpha\left[\frac{1}{6}\sigma_t(0)\langle r^2\rangle_{A'A}+
              \frac{1}{40}\sigma''_t(0)\langle r^4\rangle_{A'A}+
              \dotsb
        \right]\,,
\end{equation}
where $\langle r^n\rangle_{A'A}:=4\pi\int dr\, r^{2+n}\rho_{A'A}(r)$%
~\cite{Seltzer1969a}. Thus $F_{21}$ in the formula of PS nonlinearity
in Eq.~\eqref{Eq:PSNL} is given by $F_{21}=\sigma_2(0)/\sigma_1(0)$,
that is solely determined by the wave functions at the origin.

As for the PS, we find
\begin{equation}
X_t=(-1)^{s+1}\frac{g_ng_e}{4\pi}
    \left[\frac{\sigma_t(0)}{m^2}+3\frac{\sigma''_t(0)}{m^4}+\dotsb\right]\,,
\end{equation}
and that the leading $\sigma_t(0)$ term in the case of heavy mediator
disappears in the expression of the PS nonlinearity in Eq.~\eqref{Eq:PSNL} 
so that $\varepsilon_\text{PS}\sim O(1/m^4)$ as $m\to\infty$%
~\cite{MikamiTanakaYamamoto2017}.

In our analysis below, we numerically evaluate the integration in
$X_t$ without expanding $\sigma_t(r)$. 
The FS is also obtained by numerical integration in 
Eq.~\eqref{Eq:FS} (or \eqref{Eq:FS2}) without expansion. 
We decompose the FS into the leading and subleading contributions
as $\nu_{t,A'A}\bigl|_\text{FS}=F_t\langle r^2\rangle_{A'A}+G_t(A'-A)$.
Then, the FS nonlinearity is given by
\begin{equation}\label{Eq:FSNL}
\varepsilon_\text{FS}:=\frac{M}{m_e^2}(G_2-F_{21}G_1)\,.
\end{equation}
as in the same manner as the PS nonlinearity.
Strictly speaking, $G_t$ evaluated by subtracting the leading FS
from the total FS depends on the isotope pair $(A,A')$. 
However, our numerical result shows that the dependence is weak
and less than a few per cent among the isotope pairs 
analyzed below.

\section{\label{Sec:StatusProspect}Current status and future prospects}
We analyze the same experimental data of the singly charged calcium and 
ytterbium ions as our previous work~\cite{MikamiTanakaYamamoto2017}, 
in which NR wave functions are employed. 
The IS in the transitions of $^2\text{S}_{1/2}$ -- $^2\text{P}_{1/2}$
(397 nm) and $^2\text{D}_{3/2}$ -- $^2\text{P}_{1/2}$ (866 nm) of 
$\text{Ca}^+$ are measured with $O(100)$ kHz precision for the isotope
pairs (40, 42), (40, 44) and (40, 48)~\cite{Gebert2015a}.
The 397 nm transition is treated as the $4\text{s}_{1/2}$ -- 
$4\text{p}_{1/2}$ transition and the 866 nm as the $3\text{d}_{3/2}$ --
$4\text{p}_{1/2}$ in the single-electron approximation.
As for $\text{Yb}^+$, the IS of $^2\text{S}_{1/2}$ -- $^2\text{P}_{1/2}$
(369 nm, $6\text{s}_{1/2}$ -- $6\text{p}_{1/2}$)%
~\cite{Martensson-PendrillGoughHannaford1994a} and 
$^2\text{D}_{3/2}$ -- $^3\text{D}[3/2]_{3/2}$ 
(935 nm, $4\text{f}_{5/2}$ -- $6\text{s}_{1/2}$)%
~\cite{SugiyamaWakitaNakata2000a} are measured for (172, 170), 
(172, 174) and (172, 176). 
The experimental precision of the former transition is $O(1)$ MHz 
and that of the latter is $O(10)$ MHz.
A summary of these data and King's plots are given in
Ref.~\cite{MikamiTanakaYamamoto2017}.

\begin{table}
\caption{Nonlinearity parameters in the atomic units. 
         To convert to the natural units, multiply $\alpha^2$.
         The errors are one standard deviation.
         The values of $\varepsilon_\text{PS}$
         are those for $m=1\ \text{keV}$ and $|g_n g_e|=1\times 10^{-13}$.}
\label{TB:NL}
\centering
\begin{tabular}{cccccc}
 & $|\varepsilon_\text{exp}|$ & $|\varepsilon_\text{exp}|$
 & $|\varepsilon_\text{PS}|$  & $|\varepsilon_\text{FS}|$ 
 & $|\varepsilon_\text{FS}|$\\
 & (present) & (1 Hz) & 
 & ($\text{p}_{1/2}$) & ($\text{p}_{3/2}$)\\
\hline
$\text{Ca}^+$ & $(-2.5\pm4.1)\cdot 10^{-6}$ & $4.5\cdot 10^{-11}$ 
 & $8\cdot 10^{-11}$ & $1\cdot 10^{-11}$ & $5\cdot 10^{-13}$\\
$\text{Yb}^+$ & $(1.3\pm 1.4)\cdot 10^{-4}$ & $4.2\cdot 10^{-11}$
 & $3\cdot 10^{-9}$ & $5\cdot 10^{-8}$ & $5\cdot 10^{-10}$
\end{tabular}
\end{table}

It turns out that the data satisfy the linearity within the error.
Comparing the data with the IS formula including the nonlinearity, 
Eq.~\eqref{Eq:NL} with the PS nonlinearity parameter 
$\varepsilon_\text{PS}$ replaced by the experimental nonlinearity 
parameter $\varepsilon_\text{exp}$, we construct 
the $\chi^2$ as described in App.~E of 
Ref.~\cite{MikamiTanakaYamamoto2017} and obtain quantitative bounds
on the nonlinearity as presented in Table \ref{TB:NL}.
We also present the expected sensitivity of future experiments 
in Table \ref{TB:NL} extrapolating the experimental precision to 1 Hz. 
We note that IS measurements of 1 Hz precision are
conceivable since the IS of the $\text{Ca}^+$ $^2\text{S}_{1/2}$ -- 
$^2\text{D}_{5/2}$ (729 nm) transition is measured with a precision
better than 10 Hz~\cite{KnollmannPatelDoret2019a}.

Our prediction of the PS nonlinearity $\varepsilon_\text{PS}$
indicated in Table \ref{TB:NL} is that for a representative 
set of mediator mass and couplings in Eq.~\eqref{Eq:NewPot}, 
$m=1\ \text{keV}$ and $|g_n g_e|=1\times 10^{-13}$.
The PS nonlinearity of $\text{Ca}^+$ for this set of parameters
is the same order as the sensitivity expected in experiments of
1 Hz precision. 
For $\text{Yb}^+$, the sensitivity of 1 Hz experiments is 
sufficiently high to probe the PS nonlinearity for the same set
of parameters, provided that the FS nonlinearity is suppressed.

\begin{figure}
 \centering
 \includegraphics[width=0.5\textwidth]{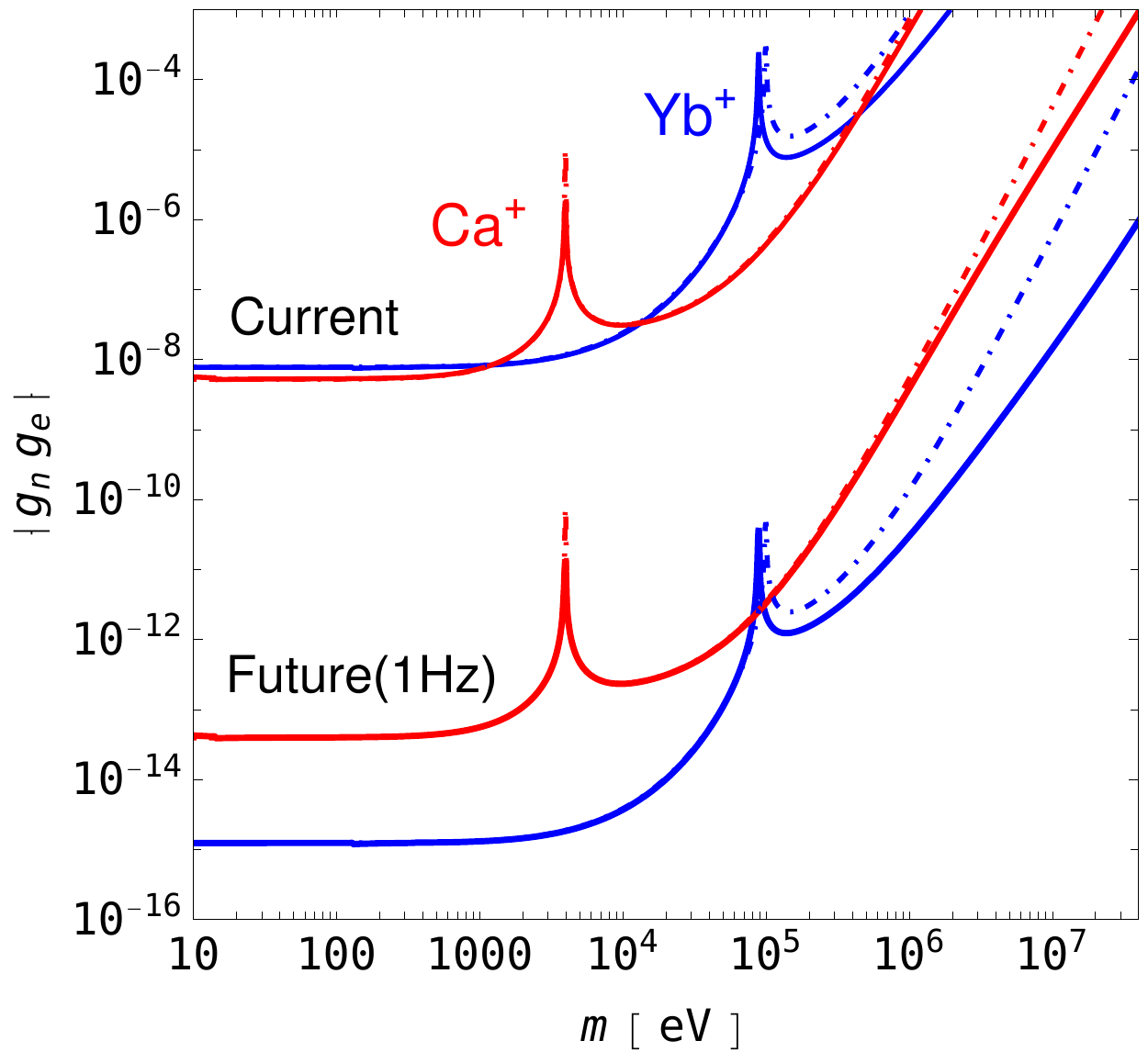}
 \caption{Constraints on mediator mass and couplings.
          The solid lines show the results of the relativistic
          calculation and the dash-dotted lines are NR.
          The red and blue lines are those of $\text{Ca}^+$ and
          $\text{Yb}^+$ respectively. 
          The upper lines represent the current constraints and 
          the lower ones are expected with 1 Hz precision.}
 \label{Fig:NRvsR}
\end{figure}

Figure \ref{Fig:NRvsR} shows the current constraints on the mass
and couplings as well as those expected with experimental data
of 1 Hz precision. 
The region above the upper lines are excluded by the present data,
while the lower lines express the expected sensitivity of future
experiments. The red (blue) lines are the case of $\text{Ca}^+$ 
($\text{Yb}^+$). The results of the relativistic calculation in 
the present work are represented by the solid lines and 
those of the NR calculation in our previous work%
~\cite{MikamiTanakaYamamoto2017} are shown by the dash-dotted 
lines for comparison. We observe that the PS nonlinearity is 
enhanced and the sensitivity is improved significantly for 
$m\gtrsim 1$ MeV in the relativistic calculation of $\text{Yb}^+$ 
compared to the NR case.
This is due to the nonvanishing electron density of 
the $6\text{p}_{1/2}$ state at the origin,
shown in Fig.~\ref{Fig:WF}.
The relativistic effect is less notable for $\text{Ca}^+$ in
the depicted mass range.

The peak structure that implies the loss of sensitivity at the
corresponding mediator mass is due to the cancellation in 
the PS electronic factor in Eq.~\eqref{Eq:PS}.
Since the peak position depends on the involved wave functions, 
it is useful for evading the sensitivity loss to combine two or
more linearity tests, e.g.~$\text{Ca}^+$ and $\text{Yb}^+$ as shown 
in Fig.~\ref{Fig:NRvsR}.

For the $\text{Ca}^+$ case, the present work employs the same 
experimental data (of the same transitions) as Ref.~\cite{Berengut2017a}.
Our single-electron calculation reasonably agrees with the more 
sophisticated one in Ref.~\cite{Berengut2017a}.
For example, the peak position of the present $\text{Ca}^+$ constraint
in Fig.~1 of Ref.~\cite{Berengut2017a} is reproduced well in 
our calculation.

\begin{figure}
 \centering
 \includegraphics[width=0.45\textwidth]{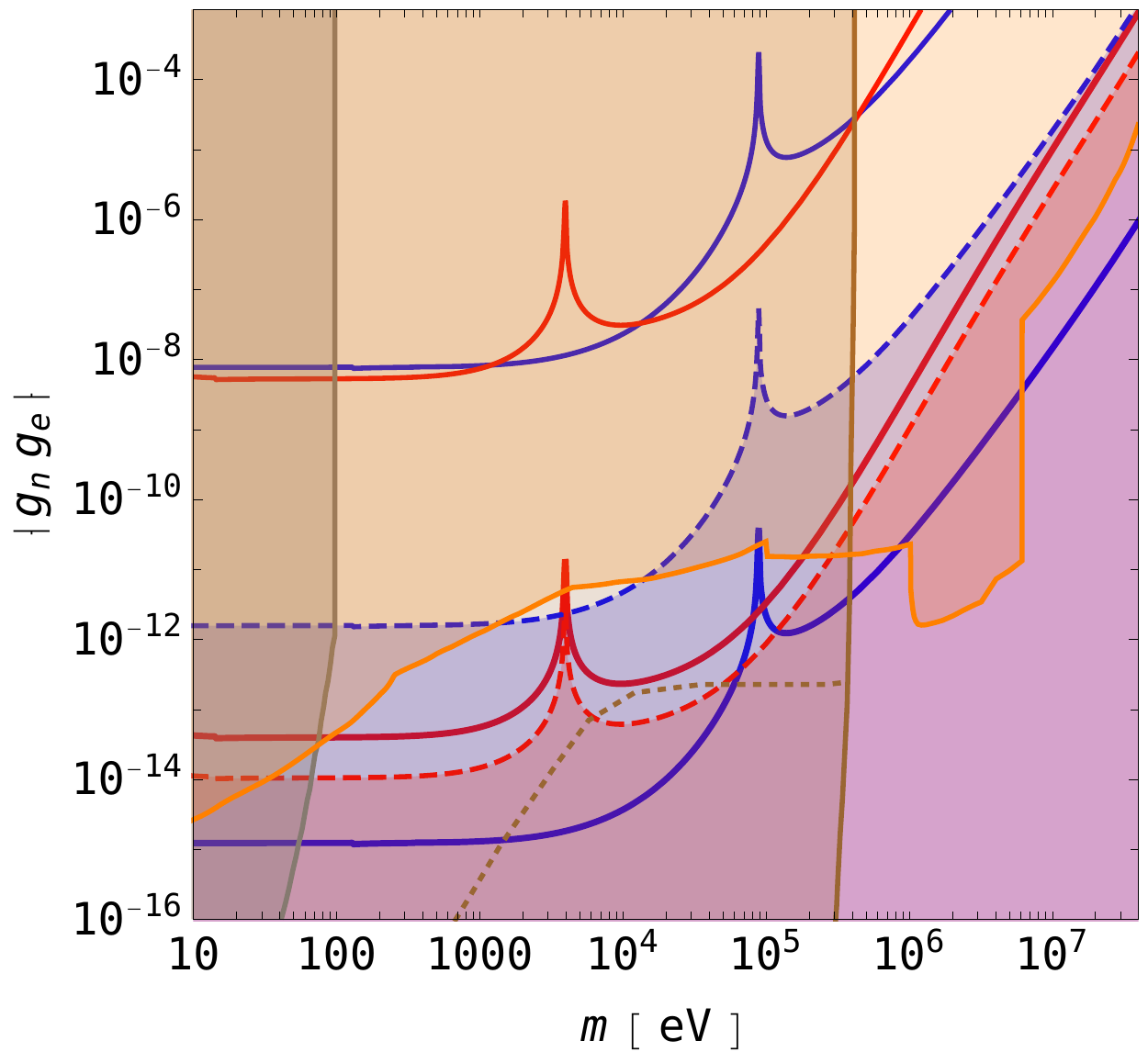}\ \ 
 \includegraphics[width=0.45\textwidth]{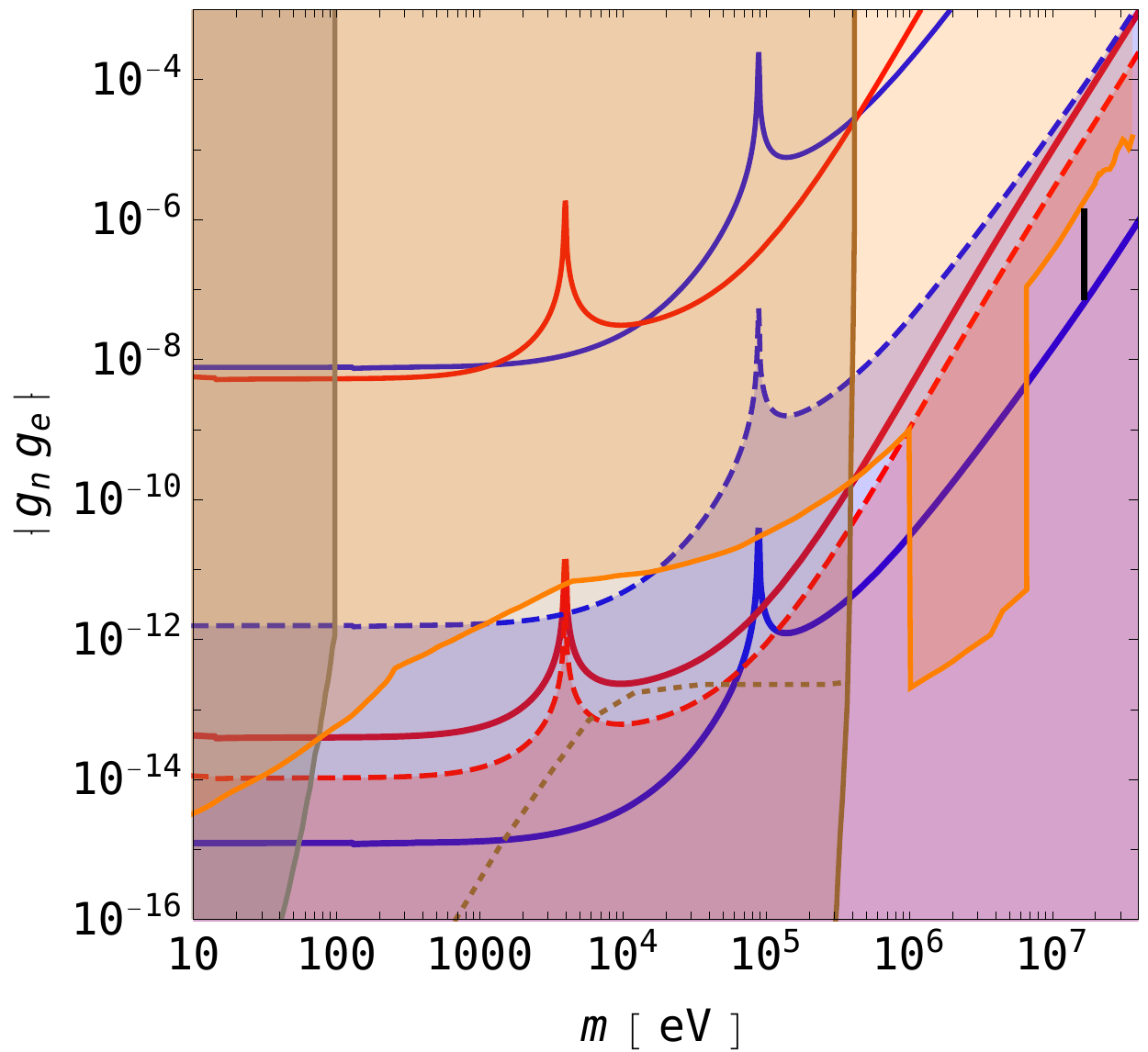}
 \caption{The constraints on the mediator mass and couplings 
          compared with other experiments/observations. The left
          (right) panel is for a scalar (vector) mediator.
          The red and blue solid lines are the same as 
          Fig.~\ref{Fig:NRvsR}.
          The grey shaded regions left of 
          the line about 100 eV
          in both panel are excluded by the fifth force 
          search~\cite{Ederth2000a,Fischbach2001a}.
          The shaded regions above the orange lines are
          the excluded regions by combining the neutron scattering data
          ~\cite{LeebSchmiedmayer1992a,Pokotilovski2006a,Nesvizhevsky2007a}
          and several constraints on $g_e$~\cite{Hanneke2008a,
          DavoudiaslLeeMarciano2012a,BabarDP2014a,Liu2016a,Andreas2012a}. 
          The left sides of the brown lines are constrained by
          the stellar cooling~\cite{AnPospelovPradler2013a},
          but there is an uncertainty above the brown dotted 
          lines~\cite{Redondo2008a}.
          The black vertical bar in the right panel shows the coupling range
          suggested by the 17 MeV Atomki anomaly~\cite{Krasznahorkay2015a,
          Feng2016a,Feng2016b,Banerjee2018a}.
          In the regions below the red ($\text{Ca}^+$)
          and blue ($\text{Yb}^+$) dashed lines,
          the FS nonlinearity dominates over the PS nonlinearity.
          See the main text for more details.}
 \label{Fig:Constraints}
\end{figure}

The current and expected constraints of IS nonlinearity are compared 
with those by other experiments and observations for each case of
a scalar and a vector mediators in Fig.~\ref{Fig:Constraints}.
The constraints by the IS are identical for both cases.
The grey shaded regions of smaller mass in both panels are excluded
by the fifth force search~\cite{Ederth2000a,Fischbach2001a}.
The shaded regions bounded from below by the orange lines 
indicate the excluded regions by the electron $g-2$ measurement%
~\cite{Hanneke2008a,DavoudiaslLeeMarciano2012a}, 
the dark photon search at BaBar~\cite{BabarDP2014a},
and the beam dump experiments~\cite{Liu2016a,Andreas2012a},
combined with the neutron scattering experiments%
~\cite{LeebSchmiedmayer1992a,Pokotilovski2006a,Nesvizhevsky2007a}.%
\footnote{The constraints by the fifth force, 
$(g-2)_e$ and the neutron scattering were first applied in 
the present context in Ref.~\cite{Berengut2017a}.}
The BaBar result is relevant in the mass region 
above 20 MeV for the vector mediator.
 The beam dump experiments are applied in the region between 100 keV
(1 MeV) and $\sim 10$ MeV for the case 
of scalar (vector) mediator
assuming no invisible decay modes of the mediator.
The electron $g-2$ is used in the rest of the mass region
of the orange lines.
The combined terrestrial bounds are not identical for the scalar and
vector mediators. 
The left side regions of the brown lines are constrained by
the stellar cooling~\cite{AnPospelovPradler2013a}.
We note that there is an uncertainty that may invalidate these
stellar constraints in the relatively strong coupling regions above
the brown dotted lines~\cite{Redondo2008a}.
The black vertical bar in the right panel indicates the coupling range
suggested by the 17 MeV Atomki anomaly in the $^8\text{Be}$ internal
conversion~\cite{Krasznahorkay2015a,Feng2016a,Feng2016b,Banerjee2018a}.

In the fifth column of Table~\ref{TB:NL}, we present
the FS nonlinearity parameter in Eq.~\eqref{Eq:FSNL} evaluated with
the relativistic wave functions.
It turns out that the FS nonlinearity is enhanced by about two orders
of magnitude compared to the NR calculation, which gives 
$|\varepsilon_\text{FS}(\text{NR})|=5\times 10^{-13}$ and $2\times 10^{-10}$
for $\text{Ca}^+$ and $\text{Yb}^+$ respectively.
This is due to the nonvanishing $\text{p}_{1/2}$ wave functions at 
the origin for both $\text{Ca}^+$ and $\text{Yb}^+$ as illustrated
in Fig.~\ref{Fig:WF} for the latter.

If $|\varepsilon_\text{PS}|$ is smaller than $|\varepsilon_\text{FS}|$, 
the experimental search of new force through the IS nonlinearity becomes
difficult.
This is the case in the shaded regions below 
the red ($\text{Ca}^+$) and blue ($\text{Yb}^+$) 
dashed lines in Fig.~\ref{Fig:Constraints}.
These dashed lines represent the expected sensitivities in experiments
of 0.3 Hz precision for $\text{Ca}^+$ and 1 kHz for $\text{Yb}^+$.
We observe that the extrapolated $\text{Ca}^+$ 
experiment of the same pair of transitions
has the potential sensitivity to probe the parameter 
space of the mass range
from $O(0.1)$ to $O(100)$ keV that has not been excluded by other 
experiments nor observations. As for $\text{Yb}^+$, 
the maximal signal of new physics is 
comparable to or smaller than the background of the FS nonlinearity
in the transition pair considered in the present work, 
although the signal sensitivity (the lower blue solid lines in 
Fig.~\ref{Fig:Constraints}) does not differ much from
that of a different pair of $\text{Yb}^+$ transitions given in 
Ref.~\cite{Berengut2017a}.

In our previous work~\cite{MikamiTanakaYamamoto2017}, we described a 
method to remove the FS nonlinearity by generalizing the linearity
relation with three or more transitions.
Here we consider an alternative way to avoid the large FS nonlinearity. 

The inclusion of two distinct states that have nonvanishing wave
functions at the origin, such as $\text{s}_{1/2}$ and $\text{p}_{1/2}$
in this work, leads to the large FS nonlinearity as argued in
Ref.~\cite{MikamiTanakaYamamoto2017}.
This is because the FS is governed by the wave functions inside
the nucleus.
Hence it is possible to suppress the FS and its nonlinearity by
replacing the $\text{p}_{1/2}$ state with a state whose wave function
vanishes at the origin, e.g.~the $\text{p}_{3/2}$ state shown in 
Fig.~\ref{Fig:WF}. 
Our estimation of the magnitudes of FS nonlinearity for the cases of
$\text{p}_{3/2}$ is presented in the last column of 
Table~\ref{TB:NL}. 
We observe that the use of $\text{p}_{3/2}$ instead of $\text{p}_{1/2}$
reduces the FS nonlinearity to the level of the NR calculation.
This is plausible because of the similarity of the $\text{p}_{3/2}$
and NR wave functions as seen in Fig.~\ref{Fig:WF}.
We note that the PS nonlinearity in the mediator mass range below
$O(0.1)$ MeV is not significantly affected by substituting 
$\text{p}_{3/2}$ for $\text{p}_{1/2}$.
It is possible in principle to employ a state other than 
$\text{p}_{3/2}$ as far as it vanishes at the origin.

\section{\label{Sec:Conclusion}Conclusion}
We have examined the implication of relativistic calculation 
in the search of new intra-atomic force using the IS nonlinearity.
The relativistic wave functions in the single-electron 
approximation are obtained by numerically solving the Dirac equation.
We have employed the effective potential described by the Thomas-Fermi
model with the Helm nuclear charge distribution.
We have evaluated the PS and FS nonlinearities with these wave functions.
The calculated nonlinearities and the current experimental bounds by
the $\text{Ca}^+$ and the $\text{Yb}^+$ IS measurements are presented in 
Table~\ref{TB:NL} as well as the expectation in future experiments. 

In Fig.~\ref{Fig:NRvsR}, the bounds on the mass and couplings of
the force mediator in the relativistic calculation are compared
with those in the NR calculation~\cite{MikamiTanakaYamamoto2017}.
We have found that the NR calculation underestimates the magnitude
of the PS nonlinearity in the mass range greater than $\sim 1$ MeV
for $\text{Yb}^+$, while the relativistic effect in the PS is less
sizable for $\text{Ca}^+$.
Our $\text{Ca}^+$ results of the single-electron approximation
in Ref.~\cite{MikamiTanakaYamamoto2017} and the present work are
consistent with the result of the same pair of $\text{Ca}^+$ transitions
in Ref.~\cite{Berengut2017a}.

The current bound and the future sensitivity of IS nonlinearity
are also compared with constraints from other experiments and
observations in Fig.~\ref{Fig:Constraints}.
It is indicated that, although the current bound is weaker than
the best other constraints, the IS nonlinearity has a potential
sensitivity to probe the unexplored parameter space in future.
This observation also agrees with Ref.~\cite{Berengut2017a}
although they considered different transitions from ours.
Future experiments of 1 Hz precision have been supposed in our
numerical calculation as an illustration.
As mentioned above, IS measurements with the precision of order 1 Hz
are likely in the near future.
When such data of two or more transitions become available, 
the sensitivity of IS nonlinearity to new physics is expected 
to be notably improved.

Moreover, it turns out that the FS nonlinearity is
significantly enhanced by the relativistic effect and
the new force search beyond the $\text{Yb}^+$ IS precision of about
1 kHz is limited for the pair of transitions employed in this work.
The $\text{Ca}^+$ case is limited at and beyond 
about 0.3 Hz, so that the problem is less serious.

The large FS nonlinearity is due to the nonvanishing value
of the $\text{p}_{1/2}$ wave function at the origin, unlike
the NR p state.
Therefore, instead of $\text{p}_{1/2}$, we have considered
the use of the $\text{p}_{3/2}$ state, whose wave function
vanishes at the origin like the NR p state. 
We have found that the FS nonlinearity is suppressed in 
the $\text{p}_{3/2}$ case as shown in Table~\ref{TB:NL}.

In conclusion, the experimental search of new physics with
the IS nonlinearity has a potential sensitivity beyond 
the existing terrestrial constraints. 
A proper selection of transitions is important as well as
improvements of experimental precision.

\section*{Acknowledgments}
The work of MT is supported in part by JSPS KAKENHI Grant Numbers 
JP 16H03993, 17H02895 and 18K03621.
The work of YY is supported in part by National Science Center
(Poland) under Grant No. 2017/26/D/ST2/00490.


\begin{thebibliography}{10}
\bibitem{Huntemann2016a}
N.~Huntemann, C.~Sanner, B.~Lipphardt, C.~Tamm, and E.~Peik, ``Single-ion
  atomic clock with
  $3\ifmmode\times\else\texttimes\fi{}{10}^{\ensuremath{-}18}$ systematic
  uncertainty,'' \href{http://dx.doi.org/10.1103/PhysRevLett.116.063001}{{\em
  Phys. Rev. Lett.} {\bfseries 116} (Feb, 2016) 063001}.
  \url{https://link.aps.org/doi/10.1103/PhysRevLett.116.063001}.

\bibitem{DelaunayOzeriPerezSoreq2016a}
C.~Delaunay, R.~Ozeri, G.~Perez, and Y.~Soreq, ``{Probing Atomic Higgs-like
  Forces at the Precision Frontier},''
  \href{http://dx.doi.org/10.1103/PhysRevD.96.093001}{{\em Phys. Rev.}
  {\bfseries D96} no.~9, (2017) 093001},
\href{http://arxiv.org/abs/1601.05087}{{\ttfamily arXiv:1601.05087 [hep-ph]}}.

\bibitem{Berengut2017a}
J.~C. Berengut {\em et~al.}, ``{Probing New Long-Range Interactions by Isotope
  Shift Spectroscopy},''
  \href{http://dx.doi.org/10.1103/PhysRevLett.120.091801}{{\em Phys. Rev.
  Lett.} {\bfseries 120} (2018) 091801},
\href{http://arxiv.org/abs/1704.05068}{{\ttfamily arXiv:1704.05068 [hep-ph]}}.

\bibitem{DelaunayFrugiueleFuchsSoreq2017a} 
  C.~Delaunay, C.~Frugiuele, E.~Fuchs and Y.~Soreq,
  ``Probing new spin-independent interactions through precision spectroscopy in atoms with few electrons,''
  \href{http://dx.doi.org/10.1103/PhysRevD.96.115002}
  {{\em Phys. Rev.} {\bfseries D96} no.~11, (2017) 115002},
\href{http://arxiv.org/abs/1709.02817}{{\ttfamily arXiv:1709.02817 [hep-ph]}}.

\bibitem{King1984a}
W.~H. King, \href{http://dx.doi.org/10.1007/978-1-4899-1786-7}{{\em {Isotope
  shifts in atomic spectra}}}.
\newblock {Springer}, first~ed., 1984.

\bibitem{King1963a}
W.~H. King, ``Comments on the article ``peculiarities of the isotope shift in
  the samarium spectrum'',''
  \href{http://dx.doi.org/10.1364/JOSA.53.000638}{{\em J. Opt. Soc. Am.}
  {\bfseries 53} no.~5, (May, 1963) 638--639}.
  \url{http://www.osapublishing.org/abstract.cfm?URI=josa-53-5-638}.

\bibitem{Wang2012a}
M.~Wang, G.~Audi, A.~Wapstra, F.~Kondev, M.~MacCormick, X.~Xu, and B.~Pfeiffer,
  ``The ame2012 atomic mass evaluation,''
  \href{http://dx.doi.org/10.1088/1674-1137/36/12/003}{{\em Chinese Physics C}
  {\bfseries 36} no.~12, (Dec, 2012) 1603--2014}.
  \url{https://doi.org/10.1088%2F1674-1137%2F36%2F12%2F003}.

\bibitem{MikamiTanakaYamamoto2017}
K.~Mikami, M.~Tanaka, and Y.~Yamamoto, ``{Probing new intra-atomic force with
  isotope shifts},''
  \href{http://dx.doi.org/10.1140/epjc/s10052-017-5467-4}{{\em Eur. Phys. J.}
  {\bfseries C77} no.~12, (2017) 896},
\href{http://arxiv.org/abs/1710.11443}{{\ttfamily arXiv:1710.11443 [hep-ph]}}.

\bibitem{Blundell1987a}
S.~A. Blundell, P.~E.~G. Baird, C.~W.~P. Palmer, D.~N. Stacey, and G.~K.
  Woodgate, ``A reformulation of the theory of field isotope shift in atoms,''
  \href{http://dx.doi.org/10.1088/0022-3700/20/15/015}{{\em Journal of Physics
  B: Atomic and Molecular Physics} {\bfseries 20} no.~15, (1987) 3663}.
  \url{http://stacks.iop.org/0022-3700/20/i=15/a=015}.

\bibitem{FlambaumGeddesViatkina2017a} 
  V.~V.~Flambaum, A.~J.~Geddes and A.~V.~Viatkina,
  ``Isotope shift, nonlinearity of King plots, and the search for new particles,''
 \href{http://dx.doi.org/10.1103/PhysRevA.97.032510}
 {{\em Phys. Rev.} {\bfseries A97} no. 3, (2018) 032510}, 
\href{http://arxiv.org/abs/1709.00600}
 {{\ttfamily arXiv:1709.00600 [physics.atom-ph]]}}.

\bibitem{Helm1956a}
R.~H. Helm, ``{Inelastic and Elastic Scattering of 187-Mev Electrons from
  Selected Even-Even Nuclei},''
\href{http://dx.doi.org/10.1103/PhysRev.104.1466}{{\em Phys. Rev.} {\bfseries
  104} (1956) 1466--1475}.

\bibitem{LewinSmith1995a}
J.~D. Lewin and P.~F. Smith, ``{Review of mathematics, numerical factors, and
  corrections for dark matter experiments based on elastic nuclear recoil},''
\href{http://dx.doi.org/10.1016/S0927-6505(96)00047-3}{{\em Astropart. Phys.}
  {\bfseries 6} (1996) 87--112}.

\bibitem{March1975a}
N.~March, {\em Self-consistent fields in atoms: Hartree and Thomas-Fermi
  atoms}.
\newblock Pergamon International Library. Pergamon Press, 1975.

\bibitem{Sakurai1967a}
J.~Sakurai, {\em Advanced Quantum Mechanics}.
\newblock Addison-Wesley, 1967.

\bibitem{SilbarGoldman2010a}
R.~R. Silbar and T.~Goldman, ``{Solving the radial Dirac equations: A Numerical
  odyssey},'' \href{http://dx.doi.org/10.1088/0143-0807/32/1/021}{{\em Eur. J.
  Phys.} {\bfseries 32} (2011) 217--233},
\href{http://arxiv.org/abs/1001.2514}{{\ttfamily arXiv:1001.2514
  [physics.comp-ph]}}.

\bibitem{Seltzer1969a}
E.~C. Seltzer, ``$k$ x-ray isotope shifts,''
  \href{http://dx.doi.org/10.1103/PhysRev.188.1916}{{\em Phys. Rev.} {\bfseries
  188} (Dec, 1969) 1916--1919}.
  \url{https://link.aps.org/doi/10.1103/PhysRev.188.1916}.

\bibitem{Gebert2015a}
F.~Gebert, Y.~Wan, F.~Wolf, C.~N. Angstmann, J.~C. Berengut, and P.~O. Schmidt,
  ``Precision isotope shift measurements in calcium ions using quantum logic
  detection schemes,''
  \href{http://dx.doi.org/10.1103/PhysRevLett.115.053003}{{\em Phys. Rev.
  Lett.} {\bfseries 115} (Jul, 2015) 053003}.
  \url{https://link.aps.org/doi/10.1103/PhysRevLett.115.053003}.

\bibitem{Martensson-PendrillGoughHannaford1994a}
A.-M. M\aa{}rtensson-Pendrill, D.~S. Gough, and P.~Hannaford, ``Isotope shifts
  and hyperfine structure in the 369.4-nm 6s-6${\mathit{p}}_{1/2}$ resonance
  line of singly ionized ytterbium,''
  \href{http://dx.doi.org/10.1103/PhysRevA.49.3351}{{\em Phys. Rev. A}
  {\bfseries 49} (May, 1994) 3351--3365}.
  \url{https://link.aps.org/doi/10.1103/PhysRevA.49.3351}.

\bibitem{SugiyamaWakitaNakata2000a}
K.~Sugiyama, A.~Wakita, and A.~Nakata,
  \href{http://dx.doi.org/10.1109/CPEM.2000.851105}{``{Diode-laser-based light
  sources for laser cooling of trapped $Yb^+$ ions},''} in {\em Conference on
  Precision Electromagnetic Measurements. Conference Digest. CPEM 2000 (Cat.
  No.00CH37031)}, pp.~509--510.
\newblock 2000.

\bibitem{KnollmannPatelDoret2019a}
F.~W. Knollmann, A.~N. Patel, and S.~C. Doret, ``Part-per-billion measurement
  of the
  $4{\phantom{\rule{0.16em}{0ex}}}^{2}{S}_{1/2}\ensuremath{\rightarrow}3{\phantom{\rule{0.16em}{0ex}}}^{2}{D}_{5/2}$
  electric-quadrupole-transition isotope shifts between
  $^{42,44,48}\mathrm{Ca}^{+}$ and $^{40}\mathrm{Ca}^{+}$,''
  \href{http://dx.doi.org/10.1103/PhysRevA.100.022514}{{\em Phys. Rev. A}
  {\bfseries 100} (Aug, 2019) 022514},
  \href{http://arxiv.org/abs/1906.04105}{{\ttfamily arXiv:1906.04105
  [physics.atom-ph]}}.
  \url{https://link.aps.org/doi/10.1103/PhysRevA.100.022514}.

\bibitem{Ederth2000a}
T.~Ederth, ``{Template-stripped gold surfaces with 0.4-nm rms roughness
  suitable for force measurements: Application to the Casimir force in the
  20-100-nm range},'' \href{http://dx.doi.org/10.1103/PhysRevA.62.062104}{{\em
  Phys. Rev.} {\bfseries A62} (2000) 062104},
\href{http://arxiv.org/abs/quant-ph/0008009}{{\ttfamily arXiv:quant-ph/0008009
  [quant-ph]}}.

\bibitem{Fischbach2001a}
E.~Fischbach, D.~E. Krause, V.~M. Mostepanenko, and M.~Novello, ``{New
  constraints on ultrashort ranged Yukawa interactions from atomic force
  microscopy},'' \href{http://dx.doi.org/10.1103/PhysRevD.64.075010}{{\em Phys.
  Rev.} {\bfseries D64} (2001) 075010},
\href{http://arxiv.org/abs/hep-ph/0106331}{{\ttfamily arXiv:hep-ph/0106331
  [hep-ph]}}.

\bibitem{Hanneke2008a}
D.~Hanneke, S.~Fogwell, and G.~Gabrielse, ``{New Measurement of the Electron
  Magnetic Moment and the Fine Structure Constant},''
  \href{http://dx.doi.org/10.1103/PhysRevLett.100.120801}{{\em Phys. Rev.
  Lett.} {\bfseries 100} (2008) 120801},
\href{http://arxiv.org/abs/0801.1134}{{\ttfamily arXiv:0801.1134
  [physics.atom-ph]}}.

\bibitem{DavoudiaslLeeMarciano2012a} 
  H.~Davoudiasl, H.~S.~Lee and W.~J.~Marciano,
  ``Dark Side of Higgs Diphoton Decays and Muon g-2,''
  \href{http://dx.doi.org/10.1103/PhysRevD.86.095009}
  {{\em Phys. Rev.} {\bfseries D86} (2012) 095009},
\href{http://arxiv.org/abs/1208.2973}{{\ttfamily arXiv:1208.2973 [hep-ph]}}.

\bibitem{BabarDP2014a}
{\bfseries BaBar} Collaboration, J.~P. Lees {\em et~al.}, ``{Search for a Dark
  Photon in $e^+e^-$ Collisions at BaBar},''
  \href{http://dx.doi.org/10.1103/PhysRevLett.113.201801}{{\em Phys. Rev.
  Lett.} {\bfseries 113} no.~20, (2014) 201801},
\href{http://arxiv.org/abs/1406.2980}{{\ttfamily arXiv:1406.2980 [hep-ex]}}.

\bibitem{Liu2016a}
Y.-S. Liu, D.~McKeen, and G.~A. Miller, ``{Electrophobic Scalar Boson and
  Muonic Puzzles},''
  \href{http://dx.doi.org/10.1103/PhysRevLett.117.101801}{{\em Phys. Rev.
  Lett.} {\bfseries 117} no.~10, (2016) 101801},
\href{http://arxiv.org/abs/1605.04612}{{\ttfamily arXiv:1605.04612 [hep-ph]}}.

\bibitem{Andreas2012a}
S.~Andreas, C.~Niebuhr, and A.~Ringwald, ``{New Limits on Hidden Photons from
  Past Electron Beam Dumps},''
  \href{http://dx.doi.org/10.1103/PhysRevD.86.095019}{{\em Phys. Rev.}
  {\bfseries D86} (2012) 095019},
\href{http://arxiv.org/abs/1209.6083}{{\ttfamily arXiv:1209.6083 [hep-ph]}}.

\bibitem{LeebSchmiedmayer1992a}
H.~Leeb and J.~Schmiedmayer, ``Constraint on hypothetical light interacting
  bosons from low-energy neutron experiments,''
  \href{http://dx.doi.org/10.1103/PhysRevLett.68.1472}{{\em Phys. Rev. Lett.}
  {\bfseries 68} (Mar, 1992) 1472--1475}.
  \url{https://link.aps.org/doi/10.1103/PhysRevLett.68.1472}.

\bibitem{Pokotilovski2006a}
{\relax Yu}.~N. Pokotilovski, ``{Constraints on new interactions from neutron
  scattering experiments},''
  \href{http://dx.doi.org/10.1134/S1063778806060020}{{\em Phys. Atom. Nucl.}
  {\bfseries 69} (2006) 924--931},
\href{http://arxiv.org/abs/hep-ph/0601157}{{\ttfamily arXiv:hep-ph/0601157
  [hep-ph]}}.

\bibitem{Nesvizhevsky2007a}
V.~V. Nesvizhevsky, G.~Pignol, and K.~V. Protasov, ``{Neutron scattering and
  extra short range interactions},''
  \href{http://dx.doi.org/10.1103/PhysRevD.77.034020}{{\em Phys. Rev.}
  {\bfseries D77} (2008) 034020},
\href{http://arxiv.org/abs/0711.2298}{{\ttfamily arXiv:0711.2298 [hep-ph]}}.

\bibitem{AnPospelovPradler2013a}
H.~An, M.~Pospelov, and J.~Pradler, ``{New stellar constraints on dark
  photons},'' \href{http://dx.doi.org/10.1016/j.physletb.2013.07.008}{{\em
  Phys. Lett.} {\bfseries B725} (2013) 190--195},
\href{http://arxiv.org/abs/1302.3884}{{\ttfamily arXiv:1302.3884 [hep-ph]}}.

\bibitem{Redondo2008a}
J.~Redondo, ``{Helioscope Bounds on Hidden Sector Photons},''
  \href{http://dx.doi.org/10.1088/1475-7516/2008/07/008}{{\em JCAP} {\bfseries
  0807} (2008) 008},
\href{http://arxiv.org/abs/0801.1527}{{\ttfamily arXiv:0801.1527 [hep-ph]}}.

\bibitem{Krasznahorkay2015a}
A.~J. Krasznahorkay {\em et~al.}, ``{Observation of Anomalous Internal Pair
  Creation in Be8 : A Possible Indication of a Light, Neutral Boson},''
  \href{http://dx.doi.org/10.1103/PhysRevLett.116.042501}{{\em Phys. Rev.
  Lett.} {\bfseries 116} no.~4, (2016) 042501},
\href{http://arxiv.org/abs/1504.01527}{{\ttfamily arXiv:1504.01527 [nucl-ex]}}.

\bibitem{Feng2016a}
J.~L. Feng, B.~Fornal, I.~Galon, S.~Gardner, J.~Smolinsky, T.~M.~P. Tait, and
  P.~Tanedo, ``{Protophobic Fifth-Force Interpretation of the Observed Anomaly
  in $^8$Be Nuclear Transitions},''
  \href{http://dx.doi.org/10.1103/PhysRevLett.117.071803}{{\em Phys. Rev.
  Lett.} {\bfseries 117} no.~7, (2016) 071803},
\href{http://arxiv.org/abs/1604.07411}{{\ttfamily arXiv:1604.07411 [hep-ph]}}.

\bibitem{Feng2016b}
J.~L. Feng, B.~Fornal, I.~Galon, S.~Gardner, J.~Smolinsky, T.~M.~P. Tait, and
  P.~Tanedo, ``{Particle physics models for the 17 MeV anomaly in beryllium
  nuclear decays},'' \href{http://dx.doi.org/10.1103/PhysRevD.95.035017}{{\em
  Phys. Rev.} {\bfseries D95} no.~3, (2017) 035017},
\href{http://arxiv.org/abs/1608.03591}{{\ttfamily arXiv:1608.03591 [hep-ph]}}.

\bibitem{Banerjee2018a} 
  D.~Banerjee {\em et~al.} [NA64 Collaboration],
  ``Search for a Hypothetical 16.7 MeV Gauge Boson and Dark Photons in the NA64 Experiment at CERN,
  \href{http://dx.doi.org/10.1103/PhysRevLett.120.231802}
  {{\em Phys. Rev. Lett.} {\bfseries 120} no.~23, (2018) 231802},
\href{http://arxiv.org/abs/1803.07748}{{\ttfamily arXiv:1803.07748 [hep-ex]}}.
\end{thebibliography}
\end{document}